\input{aipcheck}
\documentclass[
    ,final            
  ]
  {aipproc}

\layoutstyle{6x9}
\begin{document}

\title{Short-range nucleon correlations and neutrino emission by   neutron stars}


\author{Leonid Frankfurt}{
  address={Tel Aviv University, Tel Aviv, Israel}
}

\author{Mark Strikman}{
  address={Pennsylvania State University, University Park, PA}
}

\begin{abstract}
We argue that significant probability of protons with momenta 
above their Fermi surface leads for proton concentrations $p/n\ge 1/8$ to the enhancement of termally excited direct  and modified 
URCA processes within a cold neutron star, and to a nonzero  probability of 
direct
URCA processes for  small proton concentrations 
($p/n\le 1/8$). We evaluate high momentum tails of neutron, proton and electrons   distributions within a neutron star.
We expect also significantly faster neutrino cooling of hyperon stars.
\end{abstract}

\maketitle

\section{Introduction}
A normal neutron star is bound by gravitational interactions.
Global characteristics of neutron stars follow from the equations 
for the hydrostatic equilibrium in the general relativity, see \cite {Tolman}.
A neutron star can be divided into several layers: the crust, the outer 
and the inner cores. The outer core extends up to the densities 
$\rho\sim (2-3)\rho_0$,  where $\rho_0\approx 0.16 nucleon/fm^3$ 
is the nuclear matter density.  The inner core extends to the center 
of the neutron star where densities can be significantly larger 
$\sim 5-10 \rho_0$ and may contain muons, hyperons, and exotic matter.  Due to inverse $\beta$ decay, the nuclear matter dissolves into a uniform liquid composed of neutrons  at the density 
$\sim 1/2\rho_0 $, with 
\begin{equation}
x=N_{p}/N_{n} \sim 5\div 10\% , 
\end{equation}
admixture of protons and equal admixture of electrons and tiny admixture of muons, see Ref. 
\cite {Baym,PanHeis}.In the inner core the value of proton fraction is probably larger:  
$ \sim 10 \div 13\%$  \cite{Yakovlev}.  The most efficient 
neutrino cooling  reactions are due to direct  URCA processes involving neutron $\beta$ decay:
\begin{equation}
n\to p+e+\bar \nu_e,
\label{beta}
\end{equation}
and $\beta$ capture  in
\begin{equation}
e+p\to  n+\nu_e.
\label{inverse} 
\end{equation}
Thus it is worth to analyse how internucleon interactions 
influence termally excited direct URCA processes within 
 cold neutron stars. Standart cooling scenario assumes that direct URCA  processes can occur in the inner core only \cite{Lattimer2}.

In the ideal gas approximation the zero temperature neutron star   is described as the system of degenerate   neutron, proton
and electron gases with the ratio of proton and neutron densities, 
$x\ll 0.1$.  For any positive neutron density  the Pauli blocking in 
the electron and   proton  sectors guarantees stability of a neutron 
star to the neutron $\beta$-decay  cf. \cite{Weinberg, S.T}.  The number densities of protons and electrons are equal to ensure electrical neutrality of the star, so $k_F(e)=k_F(p)$.  The neutron Fermi momentum is significantly larger than the proton Fermi momentum because of the larger number of the neutrons:
\begin{equation}
x^{1/3} k_F(n)=  k_F(p). 
\label {gas}
 \end{equation}

The internucleon interaction produces nucleons with momenta 
above Fermi surfaces,   cf. Eqs.(\ref{neutron},\ref{proton}). To guarantee conservation of the electric and the baryon charges
nucleon occupation numbers  below the corresponding Fermi 
surfaces - $f_i(k,T=0)$ should be smaller than unity especially
for protons.  The nonrelativistic Schr\"{o}dinger equation with 
realistic nucleon-nucleon  interactions gives occupation numbers for protons with zero momenta $\approx 70\%$ for the nuclear matter density.  Even a larger depletion of occupation numbers  is found for protons with momenta near the  Fermi surface \cite{pholes}.

The Landau Fermi liquid approach \cite{L.L}  where momentum  distribution of quasiparticles 
coincides with the Fermi distribution for the ideal gas of fermions  is effective starting approximation for  describing strongly interacting 
liquid.  It has been explained by A.B.Migdal that nucleon distribution at zero temperature should exhibit the  Migdal jump at $k=k_F$  which justifies applicability of the Fermi step distribution at zero temperature. The value of the  Migdal jump is
equal to the renormalization factor $Z<1$  of the 
single-particle Green's  function in the nuclear matter.  The condition $Z< 1$ follows from the probability conservation \cite{Migdal} and implies that occupation numbers for nucleons with momenta 
$k < k_F$ are below one. In the limit  of small proton concentration 
Fermi surface nearly disappears since proton neighborhood is   
predominantly strongly interacting neutron medium. So the height of 
the Migdal jump for the proton distribution should decrease 
$\propto x$ for $x\to 0$. (Decrease of the Migdal jump due to a large probability of SRC has been discussed a long time ago for  the liquid $^3He$  \cite{Dygaev}. ) Thus   for a highly asymmetric mixture of  protons and neutrons the interaction tends to extend proton  momenta 
well beyond $k_F(p)$.

We show that for   the temperatures 
$T\ll$ 1MeV the presence of the high momentum proton tail  leads 
to a  different value and temperature dependence of  
URCA processes for $x\ge 1/8$,  cf. Eq. \ref{enhancement} as compared to that in  Refs.\cite{Bahcall, Friman, Lattimer} where the Fermi momentum distribution for quasiparticles was used. 
As the consequence of the presence of the high momentum proton tail the neutrino luminosity due to direct URCA processes differs from zero even for 
$x < 1/8$  i.e. in  the  region forbidden in  the ideal gas approximation for quasiparticles by the Pauli blocking and the  momentum conservation.

The electron gas within neutron star is ultrarelativistic. So the Coulomb parameter $e^2/v\ll 1$. Here $e$ is the electric charge of electron and $v=p/E\approx c$ is its velocity. Thus approximation of the free electron gas is justified. The Coulomb interaction  between protons with momenta $k\ge k_F(p)$  and electrons produces  electrons with momenta above the electron Fermi surface,  although with a tiny probability cf. Eq.(\ref{electron}).  So  the occupation probability for electrons:  
$f_e(k_e \le k_F(e),T=0)$ is slightly less than one. 

Thus the interaction produces holes in  all Fermi  seas removing the absolute Pauli blocking for the direct neutron, muon, hyperon  
$\beta$-decays.  We show however that the account of the Pauli 
blocking in the electron sector ensures stability of a neutron to the direct $\beta$ decay in the outer core of a  neutron star. 
Condition of stability may be  violated in the inner core where however use of nucleon degrees of freedom is questionable.

 If hyperon  stars exist (for the review of this subject and references see \cite{Bethe07}), neutrino luminocity due to direct $\beta$- decay may appear significantly larger than for a neutron star.

\section{The role  of the interaction}
  
High momentum nucleon component of the wave function of a 
neutron star follows directly from  the Schr\"{o}dinger equation in the limit  $k\gg k_F$ where $k_F$ is Fermi momentum. The derivation 
of the formulae is similar to that in \cite{F.S1,CFSS}.

At the leading order in $(k_F^2/k^2)$ the occupation numbers for protons and neutrons with momenta above Fermi surface are:  
\begin{equation}
f_{n}(k,T=0) \approx \left({\rho_{n}}\right)^2(\left(\frac{V_{nn}(k)}{k^2/m_N}\right)^2+ 2x\left(\frac{V_{pn}(k)}{k^2/m_N}\right)^2),
\label{neutron}
\end{equation}
and
\begin{equation}
f_{p}(k,T=0) \approx  \left({\rho_{n}}\right)^2(x^2 \left(\frac{V_{pp}(k)}{k^2/m_N}\right)^2 +
2x \left(\frac{V_{pn}(k)}{k^2/m_N}\right)^2).
\label{proton}
\end{equation}
Here $\rho_{i}$ is the density of constituent $i$ .   The factor  $V_{NN}(k)$  describes the high momentum tail of the potential of 
the $NN$ interaction.  The factor 2 in the above formulae accounts 
for the number of spin states. In the first term, this factor is cancelled due to the identity of nucleons within the pair. In the derivation of the formulae for the probability of SRCs we used
the approximation of nucleon density uniform in coordinate space
to  describe the uncorrelated part of the wave function. Thus, the 
value of the high momentum tail depends strongly on the nucleon density in the core of a neutron star.  Since $k_F(p) $ is significantly  smaller than $k_F(n)$,  the probability to find a proton with 
$k\ge k_F(p)$ for a  neutron density close to the nuclear density should be  significantly larger than in  nuclei where $x \approx 1$. 
Note also that the analysis of the recent data on SRCs in the symmetric nuclear matter found a significant  $\sim 20\%$ probability of nucleons above  the Fermi surface in nuclei which is predominantly due to $I=0$ SRCs \cite{Eli}.

The Coulomb interaction between protons from SRCs and electrons produces electrons with momenta above the electron Fermi surface.   Such electrons are ultrarelativistic  so Feynman diagrams  approach should be used to evaluate the high momentum electron component rather than the  nonrelativistic 
Schr\"{o}dinger equation. We find for  the  high momentum electron component approximate expression:
\begin{eqnarray}
\label{electron}
f_{e}(k_e\ge k_F(e),T=0) \approx (1/2)\int (d^3k_p/(2\pi))^3 f_p(k_p) 
\theta(k_p-k_F(p))\rho_{e} \cdot   & &  \\ \nonumber \cdot 
(1-f_p(k_p,T=0))\left(\frac{k_e+{3\over 4}  k_F(e) }{ \sqrt{k_e}\cdot \sqrt{{3\over 4}  k_F(e)}}\right)
\left(\frac{V_{Coulomb}(k)}{k_e-k_e^2/2m_N-{3\over 4}  k_F(e)}\right)^2. 
\end{eqnarray}
The factor $1-f_p(k_p,T=0$ is the number of proton holes which prevent Pauli blocking for the proton after interaction with the electron. Effectively, Eq.(\ref{electron})  gives  the probability for triple (e-p-n) short range correlations.  This equation can be simplified for  applications by using average quantities: 
\begin{eqnarray}
\label{electron2}
f_e(k_e\ge k_F(e),T=0)\approx 
(1/2) P_{pn} \left<H \right> \cdot
\nonumber \\
\cdot {\rho_{e}}
\left(\frac{k_e+{3\over 4}  k_F(e) }{ \sqrt{k_e}\cdot \sqrt{{3\over 4}   
k_F(e)}}\right)
\left(\frac{V_{Coulomb}(k)}{k_e-k_e^2/2m_N-{3\over 4}  k_F(e)}\right)^2. 
\end{eqnarray}
Here $P_{pn}$  is  the probability of pair nucleon correlation and $\left<H \right>  \approx P_{pn}$.

The factor $$1/2\left(\frac{\sqrt{k^2+m_e^2}+\sqrt{<k_e^2+m_e^2>}}{(k^2+m_e^2)^{1/4}<k_e^2+m_e^2>^{1/4}}\right)$$   follows  from the Lorentz transformation of the electron e.m. current, conveniently calculable from the Feynman diagrams. Here  $<k_e^2>$ is the average value of the square of electron momentum within the 
electron Fermi sea.

 \section{Impact of SRC on the direct and modified URCA processes 
at small  temperatures}
In the Landau Fermi liquid  approach at finite  temperature,  $T$    
the direct  URCA process Eqs.\ref{beta} and \ref{inverse}
is  allowed   by  the energy-momentum conservation law if the proton 
concentration exceeds $x=1/8$  \cite{Lattimer}.  The restriction 
on the proton  concentration follows from the necessity to guarantee the  momentum triangle:
\begin{equation}  
k_{F}(p)+k_{F}(e)\ge k_{F}(n),
\label{Prakash}
\end{equation}
in the absorption of electrons  by the protons.

If proton concentration is below threshold or direct URCA process 
is suppressed   due to nucleon superfluidity neutrino cooling 
proceeds through the less rapid modified URCA processes:
\begin{equation}
{n+(n,p)\to p+(n,p)+e +\bar \nu_e}, 
\label{np1}
\end{equation}
and 
\begin{equation}
{e+ p+(n,p)\to  n +(n,p)  +\nu_e},
\label{np2}
\end{equation}
in which additional nucleon enables momentum conservation.

The neutrino luminosity resulting from the  direct and modified URCA 
processes, $\epsilon_{URCA}$, was  evaluated in Ref.\cite{Lattimer}
for $x\ge 1/8$
where the Fermi distribution:
\begin{equation}
f_{i, bare}(k,T)= \frac{1}{1+\exp \frac{E_i-\mu_i}{kT}},
\end{equation}
describes
 the  Pauli blocking factors $1-f_e(k,T)$ and $1-f_p(k,T)$ in 
the final state. After integration over the phase volume of the decay products it was found:
\begin{equation}
\epsilon_{URCA}=c(kT)^6 \theta(k_F(e)+k_F(p)-k_F(n)).
\end{equation}
Here $c(x\ge 0.1)$ has been calculated in terms of the square of the electroweak coupling constant relevant for  low energy  weak interactions and the phase volume factors.

In the case of realistic NN interactions  significant fraction of protons has momenta above the  proton Fermi momentum. So 
Eq.\ref{Prakash} is satisfied  for the proton large momentum tail even 
for x smaller than 0.1.  For the sake of illustrative  
estimate we substitute in the probability of 
neutron $\beta$-decay the Pauli blocking factor  $(1-f_{p,bare}(k,T))$,  by the actual distribution of protons within the core of a neutron star.  We account for the probability of additional neutron from (p,n) correlation  by the additional factor $P_{pn}$.

To simplify the discussion we will ignore here tiny probability for  electron holes at zero temperature and  parametrize neutrino luminocity as
\begin{equation}
\epsilon_{URCA}=c (kT)^6  R,
\end{equation}
where $R$ accounts for the  role of SRC in  neutrino luminosity at small temperatures.  We find  
 \begin{eqnarray}
R\approx \kappa_{pn}^2 \left[\int [(1-f_p(k_p,T))\theta(k_F(p)-k(p))+  \right. \nonumber \\ \left.+f(k_p,T)\theta(k_p-k_F(p)] \theta(k_F(e)+k(p)-k_F(n)) d^3 k_p/(2\pi)^3 \right] \cdot \nonumber  \\
\cdot \left[\int (1-f_{p,bare}(k_p,T)) \theta(k_F(e)+k_F(p)-k_F(n))d^3 k_p/(2\pi)^3 \right]^{-1}.
\end{eqnarray}
  Here $f_{p}(k_p,T)$ is the occupation number of protons  
accounting the interaction and    $f_{p,bare}(T,k)$  is the 
Fermi distribution function  over proton momenta at nonzero temperature. The factor   $\kappa_{pn}$ is the overlapping
integral between acomponent of the wave function of the neutron star
containing pair nucleon correlation and the mean field wave function  of the star.
      For the numerical  estimate we use approximation:
$\kappa_{pn}^2=P_{pn}$. 
the first term
in the numerator of the above formulae 
and  put $T=0$ in the second term.
Using
 for the estimate $V_{NN}(k)\propto 1/k^2$ for $k\gg k_F$  
and  Eq.(\ref{proton}) to evaluate large $k$ behavior of 
$f_p \propto (1/k)^8 $  we obtain:
 \begin{equation}
R\approx \frac{(P_{pn}^2/5) \rho_n } {(m_NkT)^{3/2}},
\label{enhancement}
\end{equation}
where $P_{pn}$ is the the probability for a proton to have momentum $k \ge k_F(p)$. For  the illustration, we numerically evaluate the enhancement factor $R$  for neutron density close to $\rho_0$, 
 $x=\rho_p/\rho_n=0.1$,
and $P_{pn}=0.1$.  
So,   
\begin{equation}
R\approx 0.16 P_{pn}(MeV/kT)^{3/2}.
\end{equation}
The enhancement is significant for $kT\ll 1 \textrm{MeV}$. Remember that after one year a neutron 
star cools to the temperatures $T\le 0.01 \ \textrm{ MeV}$. 

Neutrino luminocity due to direct URCA processes decreases with decrease of $x$ but  differ from zero even for  the popular
option: $x \le 0.1$. So investigation of the neutrino luminocity of 
the  neutron stars may help to narrow down the range  of the allowed values 
of $x$.

\section{$\beta$ stability of neutron within the outer core of zero temperature neutron star}
\label{t=0}

Normal neutron star  is bound  by gravity.  Gravity does not forbid  decays of constituents of the star  if energy and momentum are conserved in the decay (the equivalence principle).

Constraints due to the energy-momentum conservation law  and the  
Pauli blocking in the electron sector work in  the  opposite directions.  Indeed, the  maximal momentum of an electron from  $\beta$-decay of a neutron with momentum $k_n$ is 
$\approx 1.19 MeV /(1-k_n/m_p)$. Hence, an electron produced in the 
neutron $\beta$ decay may fill the electron hole with momentum 
$k\approx 1 MeV/c$ only.  The dominant process  which may lead to the formation of electron holes  is the elastic interaction of an energetic proton with electrons within the free electron gas.  Energy-momentum conservation is  fulfilled   in the case of   nonrelativistic nucleons if electron in the Fermi sea kicked out  by proton has minimal energy in the range:
\begin{equation} 
E_{hole}(k)=((p-k_f+k)^2-p^2)/2m_N+k_f. 
\end{equation}
Here $p$ is the proton momentum and $k_{f}$ is the electron momentum in the final state.  Scattered electron has energy 
$k_f \ge E_F(e)$,  so it is legitimate to neglect by
the  electron mass.  Hence, the minimal energy of the hole (when electron and proton momenta are antiparallel in the initial state)  is
\begin{equation}
E_{hole} \sim (1/2\div 1/3)E_F(e),
\end{equation}
for the proton momenta around $p=0.4\div 0.5 GeV/c$ typical for SRC and decreases with increase of $p$.  Evident mismatch between energies of produced electron holes and electrons in the neutron decay guarantees that an electron from  $\beta$-decay of  a neutron can not fill an electron hole.  

In the case of  ultrarelativistic nucleon gas (inner core of a star?) 
energy-momentum conservation does not restrict energies of 
electron holes  produced in (e-p) interaction: 
\begin{equation}
E_{hole}(k)=\sqrt {(m_N^2+(p+k-k_f)^2}-\sqrt {(m_N^2+p^2)}+
\sqrt {(m_e^2+k_f^2)}
\end {equation}
In the limit $p/m_n\to \infty$ we obtain  expression for minimal 
energy of hole:
\begin{equation}
E_{hole}(k)=\sqrt {(m_e^2+k_f^2)}-k_f +k\approx k+m_e^2/k_f.
\end{equation}
However in this regime use of nucleon degrees of freedom would be questionable. We will not discuss further in this paper interesting question on the possible $\beta$  instability of neutron within the inner core of star.

Direct $\beta$ decay of muon produces electrons with momenta
up to $m_{\mu}/2$  which are not far from the  electron Fermi momentum.   So evaluation of Pauli blocking for muon, hyperon  
$\beta$-decays requires model building.

%%%

\section{Conclusions}

The reduction of the difference between neutron and proton 
momentum distributions  influences collective modes. The   most significant effect would be the tendency to suppress the superfluidity 
of protons (superconductivity)  due to  the deformation of the proton Fermi surface because of  an increase of the fraction of protons having
momenta above the Fermi surface.  Existence of SRC  will not strongly influence the  possible superfluidity of neutrons.  Note that superfluidity of neutrons will  further suppress neutron $\beta$ decay due to formation of neutron Cooper pairs near the Fermi surface.

Electrons and neutrinos in the $\beta$ decays  of hyperons, muons,     are vastly more energetic than in  neutron decay. Hence, if  hyperon 
or muon stars exist,  they  should decay  significantly more rapidly  than the neutron stars and produce larger neutrino flux.

Authors are indebted to   V.~Dmitriev, S.~Kahane, K.~Kikoin,
D.~Khmelnitskii for the discussion of  the intricate problems of the
Fermi liquid  approach, D.~Yakovlev for the illuminating
discussion of  
the physics of neutrino cooling of neutron stars, and E.~Pisetsky for the discussion of  the properties of SRC in nuclei.

\end{document}